\input harvmac
\def \pa {\Vert}

\def \ep{\epsilon}

\def \ep{\epsilon}

\def \up {\uparrow}

\def \k {\kappa} 

\def \g {\gamma}
\def \del {\partial}

\def \const {{\rm const}}
\def \ha{{\textstyle{1\over 2}}}

\def \chi {\chi}
\def \s {\sigma}
\def \p {\phi}
\def \m {\mu}
\def \n {\nu}

\def \sm {$\s$-model }

\def \inv {^{-1}}
\def \ov {\over }

\def \fourth{{{1\over 4}}}

\def \lr { \lref}
\def\np {{  Nucl. Phys. }}
\def \pl {{  Phys. Lett. }}
\def \mpl {{ Mod. Phys. Lett. }}
\def \prl {{  Phys. Rev. Lett. }}
\def \pr  {{ Phys. Rev. }}

\def \cqg {{ Class. Quant. Grav. }}

\baselineskip8pt
\Title{
\vbox
{\baselineskip 6pt{\hbox{  }}{\hbox
{Imperial/TP/95-96/70}}{\hbox{hep-th/9609212}} {\hbox{
  }}} }
{\vbox{\centerline { `No force' condition}
\vskip4pt
 \centerline {and BPS combinations of p-branes }
\vskip4pt
 \centerline { in 11 and 10 dimensions }
}}
\vskip -20 true pt
\medskip
\medskip
\centerline{   A.A. Tseytlin\footnote{$^{\star}$}{\baselineskip8pt
e-mail address: tseytlin@ic.ac.uk}\footnote{$^{\dagger}$}{\baselineskip8pt
Also at  Lebedev  Physics Institute, Moscow.} }

\smallskip\smallskip
\centerline {\it  Theoretical Physics Group, Blackett Laboratory,}
\smallskip

\centerline {\it  Imperial College,  London SW7 2BZ, U.K. }
\bigskip\bigskip
\centerline {\bf Abstract}
\medskip
\baselineskip10pt
\noindent
The condition of vanishing of static force 
on a q-brane probe in the gravitational background produced 
by another p-brane is used to give a simple
derivation of the pair-wise intersection rules which govern 
the construction of BPS combinations of branes. These rules, 
while implied also by supersymmetry considerations, 
thus have purely bosonic origin. Imposing the no-force
requirement makes possible to add  branes `one by one' to
construct composite BPS configurations (with zero binding energy) 
of 2-branes  and 5-branes in D=11 and of various  p-branes 
in D=10. The advantage of this elementary approach is its
universality, i.e. the cases of different dimensions
and different types of branes (e.g., NS-NS, R-R and 
`mixed' combinations of NS-NS and R-R branes in D=10) 
are all treated in the same way.

\medskip
\Date {September 1996}
\noblackbox
\baselineskip 14pt plus 2pt minus 2pt
\lr \dgh {A. Dabholkar, G.W. Gibbons, J. Harvey and F. Ruiz Ruiz,  \np
B340 (1990) 33;
A. Dabholkar and  J. Harvey,  \prl
63 (1989) 478.
}
\lr\mon{J.P. Gauntlett, J.A. Harvey and J.T. Liu, \np B409 (1993) 363.}
\lr\chs{C.G. Callan, J.A. Harvey and A. Strominger, 
\np { B359 } (1991)  611.}

\lr \CM{ C.G. Callan and  J.M.  Maldacena, 
PUPT-1591,  hep-th/9602043.} 
\lr\SV {A. Strominger and C. Vafa, HUTP-96-A002,  hep-th/9601029.}

\lr\MV {J.C. Breckenridge, R.C. Myers, A.W. Peet  and C. Vafa, HUTP-96-A005,  hep-th/9602065.}

\lr \CT{M. Cveti\v c and  A.A.  Tseytlin, 
\pl { B366} (1996) 95, hep-th/9510097. 
}
\lr \CTT{M. Cveti\v c and  A.A.  Tseytlin, 
IASSNS-HEP-95-102, hep-th/9512031. 
}
\lr\LW{ F. Larsen  and F. Wilczek, 
PUPT-1576,  hep-th/9511064.    }
\lr\tsem{A.A. Tseytlin, \mpl A11 (1996) 689,   hep-th/9601177.}
\lr \horts{ G.T. Horowitz and A.A. Tseytlin,  \pr { D51} (1995) 
2896, hep-th/9409021.}
\lr\khu{R. Khuri, \np B387 (1992) 315; \pl B294 (1992) 325.}
\lr\CY{M. Cveti\v c and D. Youm,
 UPR-0672-T, hep-th/9507090; UPR-0675-T, hep-th/9508058; 
  \pl { B359} (1995) 87, 
hep-th/9507160.}

\lr\ght{G.W. Gibbons, G.T. Horowitz and P.K. Townsend, \cqg 12 (1995) 297,
hep-th/9410073.}
\lr\dul{M.J. Duff and J.X. Lu, \np B416 (1994) 301, hep-th/9306052. }
\lr\hst {G.T. Horowitz and A. Strominger, hep-th/9602051.}
\lr\dull{M.J. Duff and J.X. Lu, \pl B273 (1991) 409. }
\lr \guv{R. G\"uven, \pl B276 (1992) 49. }
\lr \gups {S.S. Gupser, I.R.   Klebanov  and A.W. Peet, 
hep-th/9602135.}
\lr \dus { M.J. Duff and  K.S. Stelle, \pl B253 (1991) 113.}

\lr\hos{G.T.~Horowitz and A.~Strominger, Nucl. Phys. { B360}
(1991) 197.}
\lr\teit{R. Nepomechi, \pr D31 (1985) 1921; C. Teitelboim, \pl B167 (1986) 69.}
\lr \duf { M.J. Duff, P.S. Howe, T. Inami and K.S. Stelle, 
\pl B191 (1987) 70. }
\lr\duh {A. Dabholkar and J.A. Harvey, \prl { 63} (1989) 478;
 A. Dabholkar, G.W.   Gibbons, J.A.   Harvey  and F. Ruiz-Ruiz,
\np { B340} (1990) 33. }
\lr\mina{M.J. Duff, J.T. Liu and R. Minasian, 
\np B452 (1995) 261, hep-th/9506126.}
\lr\dvv{R. Dijkgraaf, E. Verlinde and H. Verlinde, hep-th/9603126.}
\lr\gibb{G.W. Gibbons and P.K. Townsend, \prl  71
(1993) 3754, hep-th/9307049.}
\lr\town{P.K. Townsend, hep-th/9512062.}
\lr\kap{D. Kaplan and J. Michelson, hep-th/9510053.}
\lr\hult{
C.M. Hull and P.K. Townsend, Nucl. Phys. { B438} (1995) 109;
P.K. Townsend, Phys. Lett. {B350} (1995) 184;
E. Witten, \np B443 (1995) 85; 
J.H. Schwarz,  \pl B367 (1996) 97, hep-th/9510086, hep-th/9601077;
P.K. Townsend, hep-th/9507048;
M.J. Duff, J.T. Liu and R. Minasian, 
\np B452 (1995) 261, hep-th/9506126; 
K. Becker, M. Becker and A. Strominger, Nucl. Phys. { B456} (1995) 130;
I. Bars and S. Yankielowicz, hep-th/9511098;
P. Ho{\v r}ava and E. Witten, Nucl. Phys. { B460} (1996) 506;
E. Witten, hep-th/9512219.}
\lr\beck{
K. Becker and  M. Becker, hep-th/9602071.}
\lr\aar{
O. Aharony, J. Sonnenschein and S. Yankielowicz, hep-th/9603009.}
\lr\ald{F. Aldabe, hep-th/9603183.}
\lr\ast{A. Strominger, \pl {B383} (1996) 44, 
hep-th/9512059.}
\lr \ttt{P.K. Townsend, hep-th/9512062.}
\lr \papd{G. Papadopoulos and P.K. Townsend, \pl B380 (1996) 273, hep-th/9603087.}
\lr\jch {J. Polchinski, S. Chaudhuri and C.V. Johnson, 
hep-th/9602052.}
\lr \ddd{E. Witten, hep-th/9510135;
M. Bershadsky, C. Vafa and V. Sadov, hep-th/9510225;
A. Sen, hep-th/9510229, hep-th/9511026;
C. Vafa, hep-th/9511088;
M. Douglas, hep-th/9512077. }

\lr \gig{G.W. Gibbons, M.J. Green and M.J. Perry, 
hep-th/9511080.}

\lr \dufe{M.J. Duff, S.  Ferrara, R.R. Khuri and 
J. Rahmfeld, \pl B356 (1995) 479,  hep-th/9506057.}

\lr\stp{H. L\" u, C.N. Pope, E. Sezgin and K.S. Stelle, \np B276 (1995)  669, hep-th/9508042.}
\lr \dufl { M.J. Duff and J.X. Lu, \np B354 (1991) 141. } 
\lr \pol { J. Polchinski, \prl 75 (1995) 4724,  hep-th/9510017.} 
\lr \izq { J.M. Izquierdo, N.D. Lambert, G. Papadopoulos and 
P.K. Townsend,  \np B460 (1996) 560, hep-th/9508177. }

\lr \US{M. Cveti\v c and  A.A.  Tseytlin, 
\pl {B366} (1996) 95, hep-th/9510097.  
}
\lr\TTT{I.R. Klebanov and A.A. Tseytlin, ``Intersecting M-branes as four dimensional black holes",  hep-th/9604166. }
\lr \green{M.B. Green and M. Gutperle, hep-th/9604091.}
\lr \dkl{ M.J. Duff, R. Khuri  and J.X. Lu, Phys. Rept. 259 (1995) 213, hep-th/9412184. } 
\lr \papd{
G.~Papadopoulos and P.K.~Townsend,
hep-th/9603087. }
\lr \tse{
A.A.~Tseytlin, 
\np B475 (1996) 149, hep-th/9604035.}
\lr \tsek{I.R. Klebanov and 
A.A.~Tseytlin, 
\np B475 (1996) 179, hep-th/9604166.}

\lr\gaun{J.P.~Gauntlett, D.A.~Kastor and J.~Traschen, 
hep-th/9604179.}
\lr\bal{V. Balasubramanian and {}F. Larsen, hep-th/9604189.}

\lr \town{P.K. Townsend, \pl B350 (1995) 184, hep-th/9501068.}
\lr\bergg{K. Behrndt, E. Bergshoeff and B. Janssen, hep-th/9604168.}


\lr\polch{J. Polchinski, Phys. Rev. Lett. { 75} (1995)  4724.}
\lr\berdr{ E. Bergshoeff and M. de Roo, hep-th/9603123.}

\lr\grgu{
M.B. Green and M. Gutperle, hep-th/9604091.}

\lr\berg{ E.~Bergshoeff, M.~de Roo and T.~Ort\'\i n, 
{ hep-th/9606118}.}

\lr\papnew{G. Papadopoulos and P.K. Townsend, hep-th/9609095.}
\lr\papa{G. Papadopoulos, hep-th/9604068.}
\lr\lifsh{  G. Lifschytz, hep-th/9604156.}
\lr\bachas{C. Bachas, \pl B374 (1996) 37, hep-th/9511045.}
\lr\calla{C.G. Callan, J.A. Harvey  and  A. Strominger, 
 \np B367 (1991) 60.}
\lr\call{C.G. Callan and R.R. Khuri, \pl B261 (1991) 363.}
\lr\khu{R. Khuri, \pr D48 (1993) 2947.}

\lr\schwarz{ J.H. Schwarz,  \pl B367 (1996) 97, hep-th/9510086, hep-th/9601077.}
\lr\leigh{R.G. Leigh, \mpl A4 (1989) 2767.}
\lr\doug{
M. Douglas, hep-th/9512077.}
\lr\dddo{M.R. Douglas, D. Kabat, P. Pouliot and  S.H. Shenker,
  hep-th/9608024. }

\newsec{Introduction}
Recent developments in string theory suggest the importance of better understanding  of the  structure of  composite p-brane  solutions
 $10$ and $11$ dimensions. 
 One is usually interested in stable  extreme 
configurations which preserve  some supersymmetry. 
For classical bosonic 
solutions  the  relevant condition  can be stated 
as a special property of the corresponding background fields  (e.g., special holonomy of appropriate connection), which, upon embedding of the solution in a supergravity theory, implies preservation of some amount of  supersymmetry.
It may be  useful to  try to distinguish  the condition  of residual 
supersymmetry from that of  the BPS one  since 
the latter concept is defined    already   in a  bosonic theory.\foot{The BPS condition is  obviously
 more general than that of residual  supersymmetry: 
standard supersymmetry  exists only in  space-times of certain  dimensions
while  classical bosonic BPS configurations are possible in any number
of dimensions.} 
The BPS configurations  with zero binding energy (the only 
ones which  will be considered in this paper)
are  determined by  solutions of linear Laplace equations 
 and thus satisfy the `no force' condition, allowing one to 
displace  constituents  at no cost in the energy. 

A further clarification of  the role of BPS condition in  string theory 
may  provide  clues  about   M-theory. While in string theory 
there is a  remarkable principle  of  the world-sheet conformal invariance that determines the higher-order corrections to the classical  target space effective  action, no similar principle is 
known  in 11-dimensional theory. 
 At the same time, several   backgrounds 
for which the  string \sm  has the property of exact conformal invariance  (like the fundamental string and the solitonic   5-brane)
are expressed in terms of functions which satisfy the Laplace equation
and thus obey the BPS property. The Laplace equation condition
directly follows from  the special structure 
 of  background {\it and} the condition of conformal invariance 
(the vanishing of Laplacian effectively corresponds to 
 marginality of  relevant vertex operator).
 This suggests that  some appropriate  `relaxation'
 of BPS condition may be a counterpart of conformal invariance of the string \sm in the case of  p-branes.

Below we shall demonstrate how one can determine the basic 
 features
of   composite  BPS p-brane backgrounds
by  considering a q-brane  
probe  moving   in a background produced by 
a p-brane source and findind  which relative orientations of the 
probe  and the source lead to `no-force' condition.
 Having  established 
which  combination  of a q-brane and a p-brane  preserves the 
  BPS  property one may consider  their composition     
as a source for a more  complicated  background. Next, one can add 
 another m-brane probe, orient it in a `no force' way, etc.
This simple procedure works for M-branes of 11-dimensional theory 
as well as for various (NS-NS, R-R and mixed)  p-brane 
configurations of $D=10$ type II theory.
The results are consistent  with  other approaches in $D=10$
(D-brane supersymmetry  analysis   or study of  potential between D-branes \refs{\pol,\jch,\grgu,\lifsh})
 and $D=11$  \refs{\papd,\tse,\tsek,\gaun}.

The  composite BPS solutions  which will be  considered
 in this paper 
are the ones with zero binding energy  and the dimension of common transverse space  $> 2$  (so that they are localised in transverse space 
and have well-defined  energy).
In addition, there are  two other classes of possible BPS  configurations 
which will not be discussed.
The first consists of  `no force' configurations 
with dimension of the transverse space being 1: (i) `intersection' of two NS-NS 5-branes over a string 
in $D=10$ \khu,   S-dual  intersection of two R-R  5-branes, and, more
generally,    D-brane configurations with the number of 
Dirichlet-Neumann directions  equal to 8  \refs{\jch,\grgu,\lifsh};
(ii) `intersection' of  two 5-branes  over a  line in $D=11$ \gaun.
The second class includes $\ha$-supersymmetric solutions
(sometimes obtained by applying S-duality in 10 or lower dimensions)
which can be interpreted as strongly-coupled bound states 
with non-trivial binding energy: $1_{NS} \Vert 1_R$ \schwarz,
$0\Vert 2, 1\Vert 3,$  etc.  \jch\ in $D=10$, and $2\Vert 5$ in $D=11$
\refs{\izq,\papd,\papnew}.

\newsec{$D=11$ }
We shall first consider the   $D=11$  case which is effectively 
simpler then  `dilatonic' $D=10$ one.  
Our starting point is the standard 
action for a p-brane moving in a background of 
D-dimensional metric $G_{\m\n}$  and  $(p+1)$-form field  $B_{\m_1...\m_{p+1}}$
\eqn\stan{
I_p = T_p \int d^{p+1} \s [  \sqrt { - \det \hat G_{mn} }
 + {\textstyle { 1\ov  (p+1)!} }\ep^{m_1 ...m_{p+1}}  \hat  B_{m_1...m_{p+1} } + ...] \ , }
\eqn\metr{ \hat G_{mn} = G_{\m\n}(x) \del_m x^{\m} \del_n x^\n \ , \ \ \ 
\hat B_{m_1...m_{p+1} }  = B_{\m_1...\m_{p+1}} (x) \del_{m_1} x^{\m_1}... \del_{m_{p+1}} x^{\m_{p+1}}  \ . }
It will be sufficient for our purposes to ignore other possible terms 
in the p-brane actions (e.g., 
terms involving world-volume  vector or tensor fields). 
Adding this action as a source to $S= {1\ov 2\k^2} \int d^D x \sqrt g [ R -
 {1\ov 2(p+2)!} F_{p+2}^2 ], \   F_{p+2} = d B_{p+1}$ one finds the corresponding `electric' p-brane solutions \dkl. 

 Let us choose the static gauge $ x^m = \s^m \ (m=0,1,...,p)$, so that
\eqn\yty{ \hat G_{mn} = G_{mn} (x) +  G_{ij} (x) \del_m x^{i} \del_ nx^j \ , \ \ \ } $$
\hat B_{m_1...m_{p+1} }  = B_{m_1...m_{p+1}} (x) +  B_{m_1...m_p i } (x) \del_{m_{p+1}} x^{i} + ...  \ . $$
For simplicity we  assumed that the background fields do not depend on 
$x^m=y^m$, though this is not necessary. 
We have also assumed that  $G_{mi}=0$, i.e. that the metric has a `block-diagonal' structure.  Expanding \stan\ in powers of derivatives of $x^i$ we get 
\eqn\stn{
I_p = T_p \int d^{p+1} \s [\  V (x) + 
\ha \g^{kl}_{ij}  \del_k x^{i} \del_l x^j 
 + ... ] \ , }
\eqn\pote{ V= \sqrt { - \det G_{mn} } 
 + {\textstyle { 1\ov  (p+1)!} } \ep^{m_1 ...m_{p+1}} B_{m_1...m_{p+1} } (x)  \ , }
\eqn\maa{ \g^{kl}_{ij} =  \sqrt { - \det G_{mn} } \ G^{kl} G_{ij} 
 \ , }
where $V$ is the effective static  potential. If $V$ is not constant 
there will be a force term in the corresponding equation of motion for 
the p-brane probe.  Higher order terms in the expansion give `velocity-dependent' corrections to the potential.
The `no-force'  BPS configurations  thus should  lead to  
$V=0$ or $V=\const$.  The resulting condition depends (i) on a type  of p-brane probe, (ii) on a form of background fields, and (iii) on an orientation of the probe with respect to the background (i.e. to the source
q-brane which produces it).
Similar considerations  were  used  in
some  special  cases  in \refs{\dgh,\dus,\call,\dkl}.

While the arguments below have a straightforward  generalisation to the  case of non-dilatonic  p-branes in $D$ dimensions,  we shall consider explicitly 
 the specific examples which are of most interest: 2-brane and 5-brane in 11 dimensions.

\subsec{2-brane probe in 2-brane background}
The extremal BPS  background  produced by a 2-brane source is  \dus\
($i=1,..,8 $)
\eqn\two{
d s^2_{11} =  
H^{1/3} (x) \big[ H\inv (x) (-dt^2 +  dy_1^2 + dy_2^2)
 +  dx_i dx_i \big] \ ,    }
\eqn\ff{
B_3  = - H\inv(x)  dt\wedge dy_1 \wedge dy_2 \ ,  \ \ \ \  \  \del^2 H=0 \ . }
 $H(x)$ is a harmonic function  (which 
 may  depend only on part of $x$-coordinates, e.g., as a result
of taking  a periodic array 
of  1-center solutions).   

Suppose we put a 2-brane probe in this background.
If the probe is oriented parallel to the source 2-brane, i.e.  it 
lies in the plane $y_1,y_2$, 
 then  the total effective potential 
$V$  vanishes because of the cancellation of the contributions of the 
 `Nambu'  and `Wess-Zumino' terms in \pote\ \dus\foot{Similar  conclusion is reached  if one does not fix the static gauge 
but just exands the action in powers of derivatives of $x^i$:
$V = H\inv \sqrt{ |\det (\del_m y^k \del_n y^k)|} 
- H\inv  \det (\del_m y^k) =0$.} 
\eqn\poe{ V=  [(H^{-2/3})^3]^{1/2} -  H\inv =0 \ . }
The cancellation   obviously depends on the power $1/3$ in the prefactor
in \two\ (which is thus directly related to the dimension of the brane world-volume) 
and on  the sign of the relative orientation of the source and the probe.

This implies that  adding   2-branes parallel to the source  2-brane  
and considering the field they produce one  should still get a BPS background. Indeed, the corresponding solution is given by \two,\ff\  with multicenter choice for $H$.
The `metric' $\g$ in \metr\ is flat in this case, 
\eqn\metri{
\g^{kl}_{ij} = H\inv (H^{-2/3})\inv  H^{1/3} \eta^{kl} \delta_{ij}
=  \eta^{kl} \delta_{ij} \ , }
so that corrections to the force start at {\it fourth}  order in derivatives.
This is  actually true in general for a p-brane source parallel to a  BPS 
p-brane  (for similar observations  see, e.g., 
\refs{\dkl,\bachas,\lifsh}).

What  happens if the probe is oriented orthogonally to the  $y_1,y_2$ plane,
e.g., if it lies  in the  $x_1,x_2$ plane, 
orthogonally intersecting  the source 2-brane 
 over  one point?  As follows from the structure of \two,\ff, 
in this case the effective potential \pote\ 
receives contribution only from the
first  term ($B_3$ has no components in $x^i$-directions). Counting the factors of $H$ coming from the common time direction and $x_1,x_2$ directions we find that 
they cancel out,  
\eqn\poo{
V =   [H^{-2/3} (H^{1/3})^2]^{1/2} = 1 \ ,  }
so that  again there is no force on a static probe.
In this case $\g^{kl}_{ij}$ in \metr\ is not constant,
\eqn\gaa{
\g^{kl}_{i'j'} =  G^{kl} G_{i'j'} = \g^{kl} \delta_{i'j'} \ ,  \ \ \  \ \ 
\g^{00} = - H \ , \ \ \  \g^{ab} = \delta^{ab} \ , }
where $(i',j')$ correspond to the  dimensions transverse to both source and the probe, and $ (a,b)$ label two  of $x^i$  dimensions  along which the probe is lying. Thus here  the corrections to the force
 start at {\it second } order in  the expansion in power of  velocity.

If, however, the 2-brane probe shares one spatial dimension 
with the 2-brane source,   the potential is 
\eqn\pooi{ V= [ (H^{-2/3})^2 H^{1/3}]^{1/2} = H^{-1/2} \not=\const\ ,  }  
i.e.  the configuration of the two 2-branes orthogonally 
intersecting over a line will not be in  equilibrium.

The conclusion is  that    the source  composed of the 
two  2-branes orthogonally intersecting over a point
should  produce a  static BPS background. Indeed, the corresponding  extremal 
field configuration  exists and is  an obvious generalisation of \two,\ff\
\refs{\papd,\tse} ($i=1,..,6$)
\eqn\twoo{
d s^2_{11} =  
(H_1H_2)^{1/3} (x) \big[- (H_1H_2)\inv (x) dt^2 +  H_1\inv (x)( dy_1^2 
 +  dy_2^2)  }
$$ \  + \ H_2\inv (x) ( dy_3^2 + dy_4^2)
 +  dx_i dx_i \big] \ ,    $$
\eqn\ffo{
B_3  = - H\inv_1  (x) dt\wedge dy_1 \wedge dy_2 
 -  H\inv_2 (x)  dt\wedge dy_3 \wedge dy_4\ ,  \ \ \ \  \  \del^2 H_{1,2} =0 \ . }
This $2\bot 2$ 
solution corresponds to 2-brane sources being `delocalised' in the 
internal dimensions of each other (equivalently, it can be interpreted as an 
anisotropic 4-brane \guv).  The structure of the  background  is precisely 
such that adding a 2-brane probe parallel to each of the two 2-brane planes
we get zero force, in agreement with the  possibility 
 of multicenter choices for each of 
the  two  harmonic functions $H_{1,2}$ \tse.
The metric $\g$  in this case is not  constant (if the probe is parallel to the first 2-brane then  
$\g^{00} = - H_2$,  in agreement with \gaa). This property -- 
the presence of velocity squared corrections to the force -- is generic to  backgrounds
produced by   BPS superpositions of two  or more  p-branes.

 If the 2-brane probe is oriented orthogonally to  each of the two  
  2-brane sources
we get again  $V=1$ as in \poo. Hence there  exists 
 the $2\bot 2\bot2$ solution where all of the 2-branes intersect only 
 over one point  \refs{\papd,\tse,\tsek,\gaun}, etc. 
This  provides a simple  explanation of  the `harmonic function rule' of \tse,
according to which each square of differential  of a coordinate
 of a BPS composition of branes should  be multiplied by 
the inverse power of the 
product of harmonic functions of branes 
it belongs to (relative to the common transverse space interval). 

The form  of  intersecting solutions is thus not accidental and, not unexpectedly,  is intimately connected with the structure of the corresponding p-brane actions.

\subsec{Intersections of 5-branes  with 5-branes and 2-branes}
The story in the 5-brane case  is  very similar. The basic  `magnetic' 
solution  in $D=11$ is \guv\ 
\eqn\fiv{
d s^2_{11} =  
H^{2/3} (x) \big[ H\inv (x) (-dt^2 +  dy_1^2 + ...+  dy_5^2)
 +  dx_i dx_i \big] \ ,    }
\eqn\ffi{
F_4= dB_3   = *dH  \ , \ \ \ \   \ \   \del^2 H=0 \ ,  }
where $i=1,..,5 $ and  $*dH $ is the   dual form  in $R^5_x$.
The 5-brane  probe couples to the dual `electric' potential 
$B_6$  defined by  $dB_6 = *dB_3 $, where 
the dual is taken 
 with respect to the full  11-dimensional metric (corrections due to the presence of the 
Chern-Simons $dB_3\wedge B_3\wedge B_3$  term in $D=11$ action vanish for this  solution)\foot{Other details of the  structure  of  $D=11$ 
5-brane probe action    (see \berg\ and refs. there) will not be relevant below.} 
\eqn\ffil{
B_6  = - H\inv  (x) dt\wedge dy_1 \wedge dy_2 \wedge dy_3 \wedge dy_4 \wedge dy_5 \  \ . }
The background \fiv,\ffil\ describes the `electric' 5-brane  solution of  the  `Einstein gravity + 6-form'  action. 

If the  5-brane probe is parallel to the 5-brane  plane $(y_1, ...,y_5)$
 the static potential \pote\ vanishes  due to  the 
cancellation  of the two terms in \pote\   as in  \poe, 
\eqn\poet{ V=  [(H^{-1/3})^6]^{1/2} -  H\inv =0 \ ,  }
implying the existence of the BPS configuration of multiple parallel 5-branes (which is described by  the multicenter version of \fiv,\ffi).

If the 5-brane probe  is oriented so that it orthogonally intersects 
the 5-brane source  over $n<5$ of  $y_m$  dimensions,   
then, as follows from \fiv,\yty,\pote,   
\eqn\poe{ V=  [(H^{-1/3})^{n+1} (H^{2/3})^{5-n} ]{1/2}  = H^{(3-n)/2}
       \ . }
Thus the no-force condition is realised only if the
 two 5-branes intersect over a {\it 3-space},  in agreement with 
 the suggestion  in \papd. 
The $5\bot 5$  background 
 produced by two 5-branes orthogonally intersecting over a 3-space (and delocalised in the common internal 
7-space) is given by \refs{\papd,\tse}
\eqn\wof{
d s^2_{11} =  
(H_1H_2)^{2/3} (x) \big[(H_1H_2)\inv (x) (- dt^2 +  dy_1^2 + dy_2^2 + dy_3^2)
}
$$+\  H_1\inv(x)  ( dy_4^2 +  dy_5^2) + H_2\inv(x) ( dy_6^2 +  dy_7^2)
+ dx_i dx_i \big] \ , $$
\eqn\ffi{
F_4   = *dH_1(x)  \wedge dy_4\wedge dy_5 
+ *dH_2(x)  \wedge dy_6\wedge dy_7 
\ , \ \ \ \   \ \   \del^2 H=0 \ ,  }
where $*dH$ is the dual with respect to the transverse space $R^3_x$, so that 
\eqn\ftw{
B_6  = - H_1\inv (x) dt\wedge dy_1 \wedge dy_2 \wedge dy_3 
\wedge dy_4 \wedge dy_5  
 -  H\inv_2 (x) 
 dt \wedge dy_1 \wedge dy_2 \wedge dy_3 \wedge dy_6 \wedge dy_7  \ . }
This  can be interpreted as  a background produced by 
a source  built out of  two `elementary'
 5-branes oriented along ($y_1,y_2,y_3,y_4,y_5)$ and ($y_1,y_2,y_3,y_6,y_7)$, i.e. having  a common 3-space. 

Putting  a 5-brane source  in this $5\bot 5$ background 
one learns that  the no-force condition is satisfied only if the probe is parallel to any of the two constituent 5-branes or if it intersects any of the two over a 3-space. That, in turn,  implies the existence 
of the BPS configuration of the three intersecting 5-branes, 
 each pair sharing a 3-space (and thus all three  having
 one common `string' direction) \refs{\papd,\tse}. One  may   continue the process of adding new 5-branes, getting new BPS configurations  as long as the above intersection rule is satisfied and one does not exceed the total number 10 of spatial  dimensions  \gaun.

The next natural 
step is to   consider a 2-brane probe in the 5-brane background 
\fiv,\ffi.   One learns that the potential \pote\ is constant only 
if the 2-brane shares  {\it one}  common dimension  with the 5-brane, i.e.
it intersects the $y_1,...,y_5$  hyperplane over a line and has  the second direction oriented along one of the transverse coordinates $x^i$.
In this case 
$V= [(H^{-1/3})^2 H^{2/3}]^{1/2} =1$ 
(the WZ term does not contribute  here). 
Not surprisingly, 
the same conclusion is reached by  studying the 5-brane probe in the 
background produced by the 2-brane source \two,\ff.  As far as the potential is concerned, the picture is
completely symmetric with respect to the source and the probe:
what we are discussing is just the static force between the two branes 
interacting  via  massless graviton and  3-form  field 
 exchanges (described by $R + F_4^2$ action).

As a result, there should  exist `mixed'   BPS backgrounds 
produced by intersecting 2-branes and 5-branes  under  the rule that
any $2$-brane  and $5$-brane  can intersect over a line, $2$ and $2$ can  intersect  
over a point, and $5$ and $5$ can intersect over a 3-space, namely, 
$2\bot 5, \ 2\bot 2\bot 5, \   2\bot 5\bot 5, \  2\bot 2\bot  
5\bot 5 $, etc. \refs{\tse,\tsek,\gaun}. 
This conclusion is true in the bosonic
sector of the  theory and, as clear from the above, 
 is not directly related to the embedding in the  supergravity theory 
 (residual supersymmetry  is  only a 
particular  consequence of the 
 special geometrical properties of the 
 bosonic BPS backgrounds).

\newsec{D=10}
Let us now  demonstrate   how similar simple considerations imply the rules 
of constructing  composite BPS configurations of  various branes in 
10-dimensional  type IIA,B theories. 

\subsec{p-branes of NS-NS sector}
The  NS-NS  fundamental string 
background  is encoded in  the following 
string \sm  Lagrangian  (written in  the  conformal gauge, and omitting the coupling
to the  dilaton $\p$) \refs{\dgh,\horts}
\eqn\stri{
L = H\inv (x) \del_+ u \del_-  v + \del_+ x_i \del_- x_i \ , \    \  \ \ \   \del^2 H =0    \ ,  }
where $u,v= y\mp t$, i.e.
\eqn\sti{ds^2_{10}= H\inv (x)  (- dt^2 + dy)^2 +   dx_i dx_i \ , \ \ \  
B_2 = - H\inv dt \wedge dy \  , \ \ \  e^{2\p} = H\inv \ . }
The action and  potential for a  classical  test  string moving in a  $D=10$ type II theory background
is   given  by \stan,\stn,\pote.   
If the probe  
is oriented parallel to $y$ we get zero 
potential  $V$ \pote\  and flat  metric $\g$ \maa\  \dkl\ 
(as can be seen also  
by evaluating $L$ \stri\  in the static gauge $t=\tau,\ y=\s$,\ $\del_+ u
=0,\ 
\del_- v =0$).  This  is related, via   dimensional reduction, 
  to
 analogous facts about   $D=11$ 
2-branes \dus. 
  If the string probe is oriented orthogonally to the test string
the potential is  non-constant ($V=H^{-1/2}$), i.e. this is not a BPS configuration (cf. the presence of 
a non-zero force between the $D=11$ 2-branes
 which have one common direction).

The NS-NS solitonic 5-brane background  is described by \refs{\dufl,
\chs} 
\eqn\stff{
ds^2_{10}= - dt^2 + dy_1^2  + ...+
dy_5 ^2     +  H(x)  dx_i dx_i  \ , \ \ \ \   
 dB_2 = *dH  \ ,\ \ \ e^{2\p} = H \  . }
There is obviously no  force  ($V=1$)  on the string probe 
parallel to the 5-brane 
directions $y_m$, implying the existence of
 the  composite $1 \Vert 5$ BPS 
configuration \tsem\
 (which  can be   viewed also as a   reduction of $2 \bot 5$ in $D=11$ \tse).  The `metric' $\g$ \maa\ is non-trivial ($\g^{kl}_{ij}=
H(x) \eta^{kl}\delta_{ij}$), i.e. there 
are velocity squared corrections to the force. 
Explicitly, the static-gauge action of a string probe moving in a single 
5-brane
background is ($s=2,3,4,5; \ i=1,2,3,4$) 
\eqn\ace{
I_1 = T_1 \int d^{2} \s [ 1 + \ha \del_n x^{s} \del^n x^s  + 
\ha  H(x)  \del_n x^{i} \del^n x^i 
 + ... ] \ , \ \ \ \    H= 1 + {Q_5 \ov x^2_i} \ .  }
 The same action is found for a   string probe 
 moving in the   $1\Vert 5$ bound state background   \tsem\ 
 ($ s^2_{10}= H_1\inv( - dt^2 + dy_1^2)   +  dy_sdy_s 
 +  H_5  dx_i dx_i$, etc.)  
 parallel
 to the source string: all dependence on $H_1$ cancels out.

If the string  probe is oriented orthogonally to the  5-brane
 (i.e. the string and 5-brane intersect at  one point) 
 then $V= H^{1/2}\not=\const$, 
i.e. there  is  no stable $1_{NS} \bot 5_{NS}$  configuration.

The action for a  NS-NS 5-brane  probe moving in 
$D=10$ background   \calla\ 
\eqn\stadn{
I_{5_{NS}} = T_5 \int d^{6} \s [ e^{-2\p}  \sqrt { - \det \hat G_{mn} }
 + {\textstyle { 1\ov  6!} }\ep^{m_1 ...m_{6}}  \hat  B_{m_1...m_{6} } + ...] \ , \ \ \ \ dB_6 = *dB_2 \ ,   }
differs from \stan\ by an extra dilaton  factor
in the first term.\foot{This factor  appears upon 
 dimensional reduction
if one starts with the $D=11$ 5-brane action  \stan\ and uses 
the  ansatz (relating the $D=11$ and $D=10$ space-time actions)
 $ds^2_{11} = e^{{4\ov 3} \p} dx_{11}^2 +e^{-{2\ov 3} \p} ds^2_{10},$
 where $ds^2_{10}$  corresponds to the string-frame metric.}
This factor then  appears also in the corresponding potential $V$ and metric $\g$ (cf. \pote,\maa). 

 The  potential for a 5-brane probe moving in the fundamental string background  and oriented  parallel  to the string  source
is  thus 
\eqn\peot{
V= e^{-2\p} \sqrt {-\det G_{mn}} = H  [ (H\inv)^2  ] ^{1/2} =1 \ .  }
This is of course  the same conclusion  as reached above 
for  a string  probe moving in the 5-brane background.

If we  consider  a  5-brane probe   in  a 
5-brane  background then the potential vanishes if the probe is parallel 
to the source (the contribution of the WZ term is then  non-zero).
 If the source and the probe share $n<5 $ spatial dimensions then
 \eqn\peot{
V= e^{-2\p} \sqrt {-\det G_{mn}} = H^{-1}  (H^{5-n})^{1/2} 
= H^{(3-n)/2} \ ,  }
i.e. the  static $D=10$ NS-NS 5-branes are allowed 
to intersect over 3-spaces without experiencing a force, 
just like the 5-branes of 11-dimensional theory (cf. \poe).
This is consistent with the fact that  the 
 $5\bot 5$  background in $D=11$  gives the $5_{NS}\bot 5_{NS}$  solution 
 in $D=10$ upon direct (or `periodic array') 
dimensional reduction along one of the dimensions of the common transverse 3-space.
The metric $\g$ is non-trivial in all cases except $5_{NS} \Vert 5_{NS}$.

Having  established  the rules of  no-force pair-wise  combinations   of NS-NS strings and 5-branes
one is  able to  determine 
possible  composite BPS configurations, 
e.g.,  $1_{NS}\Vert 5_{NS}\bot5_{NS}$, etc.

\subsec{R-R branes }

The p-branes supported by a R-R charge  (Dp-branes) 
 are described by 
the actions similar to \stan\ (we shall assume that $p <7$)
\eqn\stane{
I_{p_{RR}} = T_p \int d^{p+1} \s\  [\ e^{-\p}  \sqrt { - \det ( \hat G_{mn}  + \hat B_{mn}) }
 + {\textstyle { 1\ov  (p+1)!} }\ep^{m_1 ...m_{p+1}}  \hat  C_{m_1...m_{p+1} } + ...] \ , }
with the  extra dilaton factor, the 
 $\hat B_{mn}  = B_{\m\n}(x) \del_m x^{\m} \del_n x^\n $ term 
in the `Nambu'  part and the R-R  $(p+1)$-form field in the WZ part
(dots stand for other terms involving world-volume vector field \refs{\leigh,\doug}). 
$B_{\m\n}$ effectively  couples  only to  the `transverse' coordinates, so that the   analogs of the  potential $V$   and the metric $\g$ in   \pote,\maa\
are 
\eqn\poted{ V=  e^{-\p} \sqrt { - \det G_{mn}  } 
 + {\textstyle { 1\ov  (p+1)!} } \ep^{m_1 ...m_{p+1}} C_{m_1...m_{p+1} } (x)  \ ,  }
$$  \g^{kl}_{ij} =   VG^{ kl} (G_{ij}  + B_{ij})
 \ .  $$
The extremal type II background produced by a R-R charge (or a Dp-brane source) \refs{\hos}\ 
  can be  represented as   
\eqn\fef{ ds^2_{10} 
=   H^{1/2}[ H\inv (-dt^2 + dy^2_1 + ... + dy^2_p)  
+   dx_i dx_{i}]  \ , \ \ \ \  e^{2\p} = H^{(3-p)/2} \ ,   }
with $C_{p+1} = - H\inv dt\wedge dy_1 \wedge ...\wedge dy_p$
in the `electric' cases  and $dC_{7-p} = *d H $ in the
 `magnetic' ones.\foot{A simple `explanation' of the structure of the metric in \fef\ can be  given 
 by using formal  T-duality  considerations 
starting from  flat space (or 9-brane) \bergg. For $p=1,5$ this structure is
implied also by the type IIB $SL(2)$-duality relation to the  NS-NS string and 5-brane
backgrounds.}

Consider a Dq-brane probe moving in a Dp-brane  background. 
If $p=q$ and the probe and the source are parallel to each other, 
 then 
$V=H\inv - H\inv =0, \ \g^{kl}_{ij} = \eta^{kl} \delta_{ij}$
(in agreement with  similar conclusions in \refs{\bachas,\lifsh}). 
If the probe and the source share $n$  common spatial dimensions 
then the resulting potential $V$ \poted\   and the $00$-component 
of the metric $\g$ \poted\ are given by 
(the WZ term does not contribute in this case)
\eqn\ted{ V= H^{-(3-p)/4} [(H^{-1/2})^{1+n} (H^{1/2})^{q-n}]^{1/2}
         = H^{(p+q -2n -4)/4} \ , }
\eqn\aat{
 \g^{00}_{ij}   = H^{(p+q -2n -4)/4}  H^{1/2} H^{1/2} \delta_{ij} = 
  H^{(p+q -2n)/4}\ \delta_{ij} \ . }
The `no-force' condition 
\eqn\nofo{ p+q -2n =4 \ , }
is, as expected, 
 symmetric under interchanging  of the probe and the source. It is 
consistent with the condition of unbroken  $\fourth$ supersymmetry, or 
the vanishing of static potential 
(1-loop vacuum energy of the open superstring in the presence of D-branes)
derived using the open string 
theory representation of D-branes \refs{\jch, \grgu,\lifsh},  namely,  
that the number of Dirichlet-Neumann directions $(=p+q -2n)$
  should be  $4$ (or $8$),   
as well as with   T-duality
 considerations 
 \refs{\bergg,\berdr}.\foot{The case of  $p+q -2n =8$  (which 
 includes, e.g.,  
 $n=0: 5\bot 3,  4\bot 4; \ n=1: 5\bot 5$)  is special as it 
corresponds to solutions with 1-dimensional common transverse space.
The correspondence between  open string representation 
of D-branes and classical backgrounds is not clear in this case, as the 
solutions (like the ones in \refs{\khu,\gaun}) are `localised' and 
have too few isometries, 
while D-branes in the open string picture are  all simply 
related by T-duality transformations.}

The allowed configurations are thus 
\eqn\allo{\ \ \ n=0: \ 4\Vert 0, \ 3\bot 1_R, \ 2\bot 2; \ \ \ \ \ 
n=1: \  5_R \Vert 1_R,\  4\bot 2, \ 3\bot 3;   }
$$ n=2: 
\  6\Vert 2,\  5_R\bot 3,\  4\bot 4;\  \ \ \ \  \ n=3:  7\Vert 3, \ 
6\bot 4, \ 5_R\bot 5_R \ ,  $$
with the metric $\g$ being non-trivial in all cases except $p=q=n$, 
in agreement with the presence 
of velocity-squared terms in the 
force as computed in \refs{\bachas,\lifsh}.
For example, the action for  a 0-brane probe  
moving in a 4-brane background
 is thus (cf. \ace) 
\eqn\rere{
I_0 = T_0 \int d \tau [ 1 - 
\ha \del_\tau  x^{s} \del_\tau  x^s  
- \ha  H(x)  \del_\tau  x^{i} \del_\tau  x^i 
 + ... ] \ , \ \ \ \   \ \    H= 1 + {Q_4 \ov x^3_i} \ ,}
 where $x^s$ are 4 directions parallel to the 4-brane and
 $x^i$ are 5 transverse directions.
This  reproduces  the moduli space metric of 
the D0-D4 system discussed in \dddo.
As might be expected,  the same action 
is found for a 0-brane moving in the $4\Vert 0$
bound state background \refs{\tse}  
   ($ds^2_{10} = (H_0H_4)^{1/2}[ - (H_0H_4)\inv dt^2 + H_4\inv
   dy_s dy_s  + dx_i dx_i ],$ $  e^{2\p} = H_0^{3/2}H_4^{-1/2}, $
   etc.), 
     i.e. the dependence on $H_0$ cancels out.
   The action for a 
  D-string in the D5-brane background is, of course,
 equivalent to \ace.

`Adding' Dp-branes one by one 
 according to these  pair-wise  superposition  rules   
 one can explicitly construct the backgrounds produced  by several intersecting D-branes as sources, e.g.,  $2\bot 2\bot 2, 1\bot 3\bot 3,
3\bot 3\bot 3\bot 3$, etc.
\refs{\tse,\tsek,\bal}.  
By considering a test  D-brane propagating in these composite backgrounds 
one finds that
 the no-force condition is satisfied if the probe is  oriented with respect to any of the constituent branes according to \nofo,\allo.

\subsec{`Mixed' configurations of NS-NS and R-R branes} 
Let us now  study the   Dp-brane probes moving in the NS-NS 
string \sti\  and 5-brane \stff\ backgrounds, and vice versa, 
the elementary string and  NS-NS 5-brane probes moving in the   R-R
 p-brane background \fef. The resulting no-force conditions 
should of course be  invariant under  interchange 
of the probe and the source. This provides a  useful 
consistency check.

If the Dp-brane probe  is  placed   in the fundamental 
 string background so that the string source is
 orthogonal to the  Dp-brane hyperplane  
 then  the resulting potential  \poted\ is  
 constant  due to the cancellation between the dilaton  factor and 
the factor of the time component of the metric in \sti,
$V = H^{1/2} H^{-1/2} =1$. 
  Thus  there  should exist stable  configurations 
 where a  fundamental string  orthogonally 
intersects a D-brane over a point, i.e. 
 $1 \Vert 0$,  $1_{NS} \bot 1_R$, $1_{NS} \bot 3$, $1_{NS} \bot 5_R$,
  etc.\foot{This is  an   analogue,   for the static classical 
BPS backgrounds,  of  the expectation that the  
fundamental strings can `end' on D-branes 
  \refs{\ast,\jch}.}

If the  Dp-brane with $p>0$ is parallel to the NS-NS string  source then 
 $V$ \poted\ is nontrivial 
\eqn\oted{ V= 
  H^{1/2}  (H^{-2})^{1/2} = H^{-1/2} \ ,  }
i.e.  `$1_{NS} \Vert$ Dp-brane' configurations do not satisfy the  `no-force' condition.  Note in this connection  that the  BPS bound state 
 $1_{NS} \Vert 1_R$  \refs{\schwarz}  has a non-vanishing  binding energy
(similar  bound states are  $2\Vert 5$ in $D=11$ \refs{\izq}
and $2\Vert 0, 3\Vert 1, 4\Vert 2$, etc. in $D=10$  \refs{\jch,\papa,\papnew}).
Such   configurations (which are $\ha$-supersymmetric)  have a different 
 (`strong-coupling') nature 
than  the  BPS bound states with zero binding energy 
discussed  in this paper (which preserve  $\fourth$ or 
less of maximal supersymmetry).


For a Dp-brane probe placed  in the solitonic 5-brane 
background \stff\  so that   the probe and the source  are  parallel 
to  $n$ common  spatial dimensions
 (with $p-n \leq 4$ to fit into the $D=10$ space-time)
 we find that $V$ in \poted\ is given by  
\eqn\pooyt{
V = H^{-1/2} (H^{p-n})^{1/2}  = H^{(p-n-1)/2} \ . }
The  stable  configurations  must have $ p-n=1$, i.e. all but one dimension of the $D$-brane should be parallel to the NS-NS 5-brane, i.e. 
\eqn\five{
n=0: \ 5_{NS} \bot 1_R ; \ \ \
n=1: \   5_{NS} \bot 2; \ \ \    n=2: \ 5_{NS} \bot 3; }
$$  n=3: \ 5_{NS} \bot 4; \ \ \  n=4: \ 5_{NS} \bot 5_R ;
\ \ \  n=5: \ 5_{NS} \Vert 6  \ . $$
These  results  are consistent with   some  
 of the already obtained above:
$5_{NS} \bot 1_R$  is related to $5_{R} \bot 1_{NS}$  and 
$5_{NS} \bot 3$ to  $5_R \bot 3$
by  S-duality of type IIB theory, $ 5_{NS} \bot 2$ of type IIA theory 
is the  direct dimensional reduction  of $5\bot 2$ in 11 dimensions, 
 $5_{NS} \bot 4$ is the reduction of 11-dimensional  $5\bot 5$.

Equivalent  conclusions  are, indeed,  
 reached  by considering  the NS-NS 
string  or  5-brane probes moving in the Dp-brane background  \fef.
The  test  string  `feels' only the metric in \fef\  while  the test 5-brane 
is directly coupled only 
to the metric and the dilaton  components of the Dp-brane background. 
 If  the string is  oriented orthogonally to  the Dp-brane, 
then $V= [  H^{-1/2} H^{1/2}]^{1/2} =1$, while if it is parallel to the Dp-brane
then $V= [(H^{-1/2})^2 ]^{1/2} =H^{-1/2}$, in agreement with \oted.
 A test NS-NS 5-brane   parallel to  $n$  spatial 
dimensions of  the Dp-brane  has  
the potential \peot\  
\eqn\yty{V=   H^{- (3-p)/2} 
[(H^{-1/2})^{n+1} (H^{1/2})^{5-n}]^{1/2} = H^{(p-n-1)/2} \ , }
i.e.  the same as in \pooyt.

\newsec{Concluding remarks}
The aim of the present paper was to demonstrate how one can 
determine the structure of composite BPS configurations of p-branes using
as an input only the knowledge of 
 (i)  the 
background  fields produced by a single brane source 
and (ii)  the basic terms in the  action for a  p-brane probe. 
The  form of the background
encodes the information about relevant massless fields  and their couplings to the brane.

The resulting rules of superposing  branes 
 in $D=10$ are,  essentially,   implied by   the rules in $D=11$.
The 
 explicitly known  BPS  
 configurations in $D=11$ 
(`boosted'  2-brane $2\up$, `boosted' 5-brane $5\up$,
intersections $2\bot 2, ..., 2\bot 5\bot 5$, etc., 
intersections with `boosts' $2\bot 5\up$,  etc. \refs{\tse,\tsek})
may be viewed as an economic way   of describing, via dimensional reduction,  various   `mixed'   backgrounds 
in $D=10$. If the $D=11$  configuration satisfies 
the `no-force' rules, the same will be true for $D=10$ configurations which are related to it by dimensional reduction.

To illustrate this, let us consider some examples, following \refs{\tse,\tsek}.
$2\up$ leads to  two  IIA  solutions in $D=10$:
 (i) reducing along the  boost direction we get $1_{NS} \pa 0$, i.e. a 
 bound state of a  fundamental string   and a 
0-brane\foot{The knowledge of explicit solution for a  0-brane lying on a 
closed 
string   ($1\pa 0$)  is interesting since this is the only case 
where an  `intersecting' configuration  formally  looks  like an  `ending'
 configuration  (or like  an  open string  wound around the 
compact direction, with both ends attached to the 0-brane). 
Note, however, that  starting with $2\up$ in $D=11$ we are
getting a `smeared' solution, i.e. the 
0-brane is not localised on the  fundamental string
(the
 harmonic functions for the string and the 0-brane  both depend only on 
8  transverse directions).} 
 (T-dual to the boosted R-R string of type IIB,  $1_{RR}\up$);  
(ii) reducing  along the other internal direction gives 
$1_{NS}\up$ (T-duality  gives back the  boosted NS-NS string, now as a IIB  solution).
Note that the two different reductions of $D=11$ solution are related
by T- and S- dualities in $D=10$. Similar relations  are often  true in other cases.
{}For $5\up$ one finds: (i) reducing along the  boost direction we get $4\pa 0$, i.e. a  bound state of a  4-brane   and a 0-brane 
(T-dual to $3\bot 1_R$ type IIB solution);   
(ii) reducing  along any  other internal direction gives 
$4\up$; T-duality then   gives  either  $(3\up)_1$ (the subscript indicates
the  number of extra 
compact isometric transverse dimensions, the total  number of which remains the same in the process of   reduction and T-duality) 
or $3\bot 1_{NS}$. 
$2\bot 2 $ in $D=11$ reduces to $2\bot 1$.

$2\bot 5 $ 
has the following  reductions to  $D=10$:
(i) $2\bot 4$ (T-dual to  $1_{R}\pa 5_{R}$, 
$3\bot 3$, $(1_{R}\bot 3)_1$);\ (ii)  $1\bot 4$ (T-dual to   $5_{R}\up$, 
$(1_{NS}\bot 3)_1$); \ (iii) $1\pa 5$ (T-dual to  $5_{NS}\up$ and  
$1_{NS}\pa 5_{NS}$  as a  IIB solution, related by  S-duality to $1_{R}\pa 5_{R}$).
 $5\bot 5 $  reduces to: 
(i) $5\bot 4$ (T-dual to  
$(5_{NS}\pa 3)_1$ 
 and 
$5_{NS}\bot 3$);\ \   (ii)  $4\bot 4$ (T-dual to $(3\bot 3)_1$  and $3\bot 5_R$).  

Similar  analysis can be repeated for
 $2\bot2\bot2, \  2\bot 2\bot 5$, etc.
As a result, starting with one configuration in $D=11$ and reducing it in different ways  
one gets various  BPS   configurations in $D=10$ 
(with some being   related by  T-duality and  type IIB S-duality), all of which, remarkably,  having a common origin in $D=11$ theory.

\newsec{ Acknowledgements}
I am grateful to I.R. Klebanov for useful discussions of  related  issues.
This work was partially  supported  by  PPARC, the European Commission TMR programme ERBFMRX-CT96-0045
 and NATO grant CRG 940870.

\vfill\eject
\listrefs
\end